# Transferring the attoclock technique to velocity map imaging

**Matthias Weger, Jochen Maurer,[*] André Ludwig, Lukas Gallmann, and Ursula Keller**

*Department of Physics, ETH Zurich, Wolfgang-Pauli-Str. 16, 8093 Zurich, Switzerland*
*[*]jocmaure@phys.ethz.ch*

**Abstract:** Attosecond angular streaking measurements have revealed deep insights into the timing of tunnel ionization processes of atoms in intense laser fields. So far experiments of this type have been performed only with a cold-target recoil-ion momentum spectrometer (COLTRIMS). Here, we present a way to apply attosecond angular streaking experiments to a velocity map imaging spectrometer (VMIS) with few-cycle pulses at a repetition rate of 10 kHz and a high ionization yield per pulse. Three-dimensional photoelectron momentum distributions from strong-field ionization of helium with an elliptically polarized, sub-10-fs pulse were retrieved by tomographic reconstruction from the momentum space electron images and used for the analysis in the polarization plane.

## 1. Introduction

In recent years, a number of experiments using the attoclock technique [1, 2] have addressed fundamental questions in ultrafast science, including tunneling delay time [2, 3], tunnel geometry [3] release times [4] and correlations [5] in sequential double ionization [4]. This technique uses the angular streaking of the electron and the ion after the ionization of the target to obtain time information about ultrafast processes in atoms and molecules on the attosecond scale. So far all attoclock experiments have been performed with a cold target recoil ion momentum spectrometer (COLTRIMS) [6], a coincidence detector that offers the possibility to record the full kinematic information of each ionization event. The drawbacks of COLTRIMS measurements are the restriction to a low count rate originating from the coincidence detection and the rather low momentum resolution compared to other momentum imaging techniques. Most recently, moving from 1-kHz to 10-kHz pulse repetition rates for COLTRIMS measurements the attoclock method was used to resolve an angular offset that suggests a real and probabilistic tunneling time [7].

A different charged-particle momentum imaging technique that is widely used in atomic, molecular and optical science is velocity map imaging (VMI) [8, 9]. It offers reduced technical complexity, higher momentum resolution and a significantly higher ionization yield per pulse compared to a COLTRIMS setup. Because the ionization rate and thus the signal strength are only restricted by the onset of space charge effects, the setup allows few hundred ionization events per laser pulse. A velocity map imaging spectrometer (VMIS) images only two-dimensional projections of the charged particle distribution, typically in the set of planes that are perpendicular to the polarization plane. The polarization plane, i.e. the plane perpendicular to the beam propagation direction, is, however, the plane of interest in attoclock experiments. The demand to exploit the momentum distribution in three dimensions and in particular the polarization plane is fulfilled by the combination of VMI and tomographic reconstruction that has been used by various groups in several applications in atomic, molecular and optical science [10-13]. The method can be used for momentum distributions with arbitrary symmetry. A different and more common method to retrieve three-dimensional charged particle distributions from two-dimensional projections is the application of a reconstruction algorithm that results in an Abel-inversion, like e.g. BASEX [14]. Algorithms that create an Abel inversion are however restricted to momentum distributions with cylindrical symmetry.

In this article, we present a way to implement the attoclock with a VMIS with high count rate and thus short data acquisition time and high momentum resolution employing few-cycle pulses. The article is structured as follows. In section 2, we give an introduction to the attoclock method. The experimental details are described in section 3 and the data processing, analysis and results are described in section 4. Furthermore, the results are compared with COLTRIMS measurements. The conclusions drawn from the work in this article follow in section 5.

## 2. Principle of the attoclock

The attoclock technique has been used to experimentally resolve the single electron tunneling delay time for the first time [2, 7]. How to measure a tunneling time has been a long ongoing debate because time is not an operator in quantum mechanics. For example, very often the exact start of the tunnel process is also not well defined.

The attoclock extracts time from two independent measurements based on two observables that can be measured independently without mutual restrictions (i.e. the two operators of the observables commutate). Observable 1 is the polarization axis of the elliptically polarized light and observable 2 is the electron momentum vector. The first observable determines time zero and the second measures time the same way as time is measured with a normal "stop watch". In the polarization plane the fast rotating electric field vector of the laser pulse is the "watch hand" and time is fully defined with the angular coordinate of the electric field vector. The attoclock has the unique advantage that both the clock and the start of the tunneling process is clearly defined and can be measured. The clock is based on the close-to-circularly polarized laser light with a clearly defined rotation period. The start of the tunneling process is "time zero" and is the exact moment in time when the electric field inside the short laser pulse reaches its maximum where we observe the highest tunnel probability. This maximum electric field is along the main axis of the polarization ellipse. Thus both the clock and the start of the tunneling process (i.e. time zero) is experimentally fully accessible with a simple independent laser polarization measurement and a standard pulse characterization using for example the SPIDER technique [15]. The tunneling delay time is reconstructed from the angular coordinate of the final electron momentum vector, typically measured with the COLTRIMS apparatus. This measurement is based on the "peak search in electron counts" as a function of the angle offset of time zero and does not need to resolve the tunneling probability distribution which of course is still present.

The final electron momentum vector from strong-field-ionized helium can be well explained by a semi-classical model [16] with two steps: the first step is the electron tunneling only described in quantum mechanics and the second step is the classical electron propagation in the laser field and ion potential. Models that neglect the interaction between the electron and the ion, predict a final angular coordinate (i.e. the streaking angle) of 90 degrees between the major axis of the polarization ellipse and the peak of the photoelectron angular distribution assuming no tunneling delay time [17]. In helium the ion potential is very simple and combined with the laser field we have a well-defined tunnel geometry [3] from which we can resolve and reconstruct the electron tunneling delay in this two-step model [7].

Therefore, the key elements of an attoclock measurement are elliptically polarized few-cycle pulses, accurate polarization control and a detector that is capable of imaging three dimensional momentum distributions of charged particles (electrons or ions).

## 3. Experimental details

*3.1 Velocity map imaging spectrometer*

For the experiments we designed and built a VMIS that meets the requirement of high ionization rate and offers a back-focusing geometry. A schematic of the setup is shown in Fig. 1. In our case a high count rate is of particular interest since the tomographic reconstruction requires good statistics to allow for extraction of distributions that are free of artifacts and low in noise. In order to achieve high count rates and to keep the data acquisition time as low as possible, it is desirable to have a high pulse repetition rate as well as a high number of ionized electrons per pulse. A high number of ionization events per pulse is achieved by the high target density in the focal volume in combination with an adequate laser peak intensity. Typical data acquisition times range from few minutes up to two hours, depending on the laser intensity. The acquisition time for the data shown in Fig. 6 was 40 minutes. The high target density in the interaction volume is accomplished with a rather unconventional design with the gas nozzle integrated into the repeller plate [18]. The use of few-cycle pulses dictates the back-focusing geometry. A silver mirror with focal length of 50 mm was used to create a tight focus with a short Rayleigh length. Since the special repeller design with the integrated gas nozzle leads to a decreased region with optimal electron imaging compared to setups with three open electrodes, it is beneficial to have a short Rayleigh length to keep the volume where the electrons are created as small as possible. From simulations we know that a resolution of $\Delta p / p \leq 0.01$ is achieved for electrons that are created 1 mm apart from each other in the interaction region. Thus, the resolution in momenta for electrons in the VMIS is better than the resolution of 0.05-0.2 a.u. for ions and around 0.1 a.u. for electrons in the COLTRIMS. Since the Rayleigh length is smaller than 1 mm for our beam and the chosen focal length of 50 mm, we make sure that the photoelectrons are created within the imaging volume. Although the back-focusing mirror poses the complexity of having an additional potential surface between the electrodes, we ensured that the mirror is not a problem for high-quality electron momentum imaging. This was validated with the ring structure from above threshold ionization [19] that is shown in Fig. 2. The absence of any molecular contaminant was confirmed by the absence of any Coulomb explosion channels [20] when the spectrometer voltages were switched to ion imaging. An example of an ion image recorded with linear polarization is shown in Fig. 3. This is of course important to make sure that our reconstruction is based on data from helium ionization because the VMIS is not based on coincidence detection. The background pressure was on the order of $10^{-10}$ mbar.

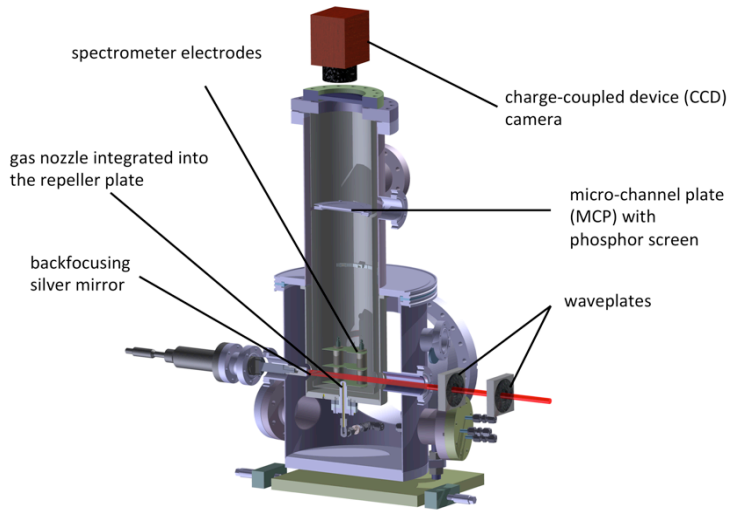

Fig. 1. Schematic of a cut through the VMIS. The incoming laser beam is marked in red.

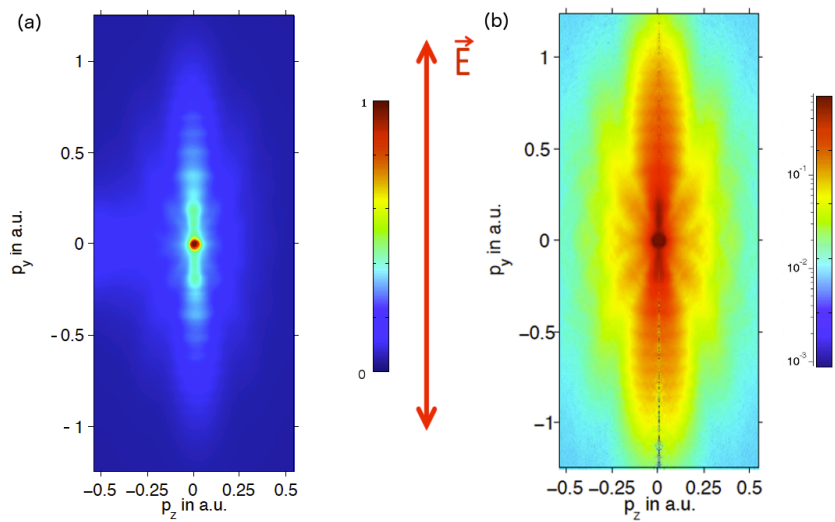

Fig. 2. Electron momentum images recorded with linear light. (a) An original momentum image as it was recorded plotted linearly scaled. (b) A cut through the three-dimensional electron momentum distribution of the image in (a) after reconstruction with BASEX and plotted on a logarithmic scale. The ATI-structure confirms the high quality of the momentum imaging.

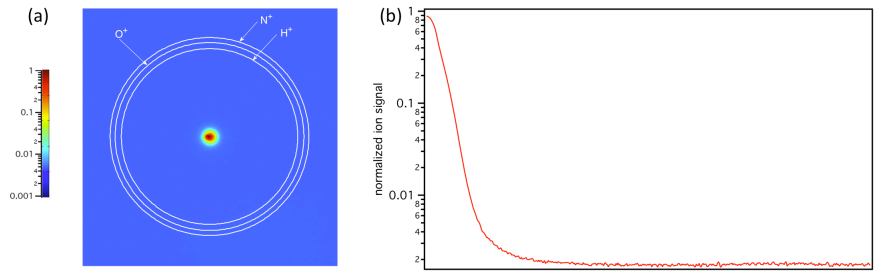

Fig. 3. (a) Ion image on logarithmic scale recorded with linear polarization at an intensity of ~0.2 PW/cm$^2$. The circles indicate the expected momenta of ion fragments H$^+$, O$^+$, N$^+$ from Coulomb explosion of H$_2$, O$_2$ and N$_2$. (b) Radial distribution of the ions from the image in Fig. (a) on a logarithmic scale.

*3.2 Optical setup*

A commercial Ti:sapphire laser system with a multipass amplifier (Femtolasers, Femtopower Compact Pro HR CEP) that delivers pulse energies of up to 750 µJ at a central wavelength of 797 nm and a pulse length of 30 fs at a repetition rate of 10 kHz was used for the experiment. The output from the amplifier was spectrally broadened in a hollow-core fiber filled with approximately 1 bar of neon and recompressed by a chirped-mirror compressor (Femtolasers, GSM 014). The pulse length was minimized by inserting an appropriate amount of fused silica into the beam path. The pulse characterization with a SPIDER resulted in a measured pulse length of 6.1 fs at a central wavelength of 735 nm. Before the pulse entered the vacuum chamber through the entrance window it passed a polarizer, a quarter quarter-wave plate (QWP) and a half-wave plate (HWP). The polarizer (Newport polarcor 05P109AR.16) ensured a clean linear polarization state of the beam before the pulse passes through the QWP. The desired ellipticity of 0.87 was induced by the quarter-wave plate (B.Halle Nachfl. GmbH RAC 5.4.10L). For the acquisition of the electron images under a large number of angles, the polarization was rotated by a superachromatic HWP (B.Halle Nachfl. GmbH RSO 2.2.10) that was mounted on a motorized rotational stage with an electronic readout with a resolution of better than 0.1 degree. The angular orientation of the waveplate is read out from an encoder on the waveplate mount. This method to determine and control the orientation of the waveplate removes any susceptibility to backlash of the rotational mount. The polarization control is discussed in more detail in section 3.4.

*3.3 Polarization characterization*

The polarization was characterized with a beam-splitting cube polarizer followed by a power meter that was placed between the optical setup and the entrance window of the VMIS. The transmission through the polarizer was measured as a function of the orientation of the major polarization axis. The result gives a calibration of the orientation of the major polarization axis with respect to the laboratory frame. The result was fitted with a $\cos^2$ function. The ellipticity and the orientation of the HWP for which the major polarization axis lies horizontally in the laboratory frame were extracted from the fit. A reconstructed electron momentum distribution from linearly polarized pulses was used to cross-check the result from the polarimetry.

*3.4 Polarization control*

An excellent control over the polarization of the laser pulse is crucial for the attoclock experiments. In our case, an accuracy of better than a few degrees is required. The use of few-cycle laser pulses with a large bandwidth poses further challenges on the control of the polarization. Dispersion is a concern with regard to the wavelength dependence of the retardation as well as the temporal spreading of the short pulses that is induced by the retarding material. The dependence of the streaking angle on the ellipticity leads to inaccurate results when the experimentally obtained streaking angle is compared with the theoretical value assuming a wrong ellipticity. Furthermore, the orientation of the major polarization axis with respect to the detector needs to be well characterized for accurate results. In order to quantify the error that is induced by the waveplates, we numerically propagated the pulse through these plates based on the specifications obtained from the manufacturer and the measured spectrum of our laser pulse [17].

Polarimetry measurements were done for two different types of HWPs, an achromatic and a superachromatic HWP from B.Halle Nachfl. GmbH. The results shown in Fig. 4 indicate that it is necessary to use the superachromatic HWP for our experiment in order to achieve a sufficiently well-defined polarization state of our few-cycle pulses. This choice becomes obvious from our simulation shown in Fig. 5 of the ellipticity as well as the polarizer transmission as a function of the angle of the HWP. For the large bandwidths in our experiment, the simulations show the desired behavior only for the superachromatic waveplate, i.e. the orientation of the major polarization axis depends linearly on the orientation of the HWP and the ellipticity is independent of the orientation of the HWP within a deviation of 0.08% in ellipticity.

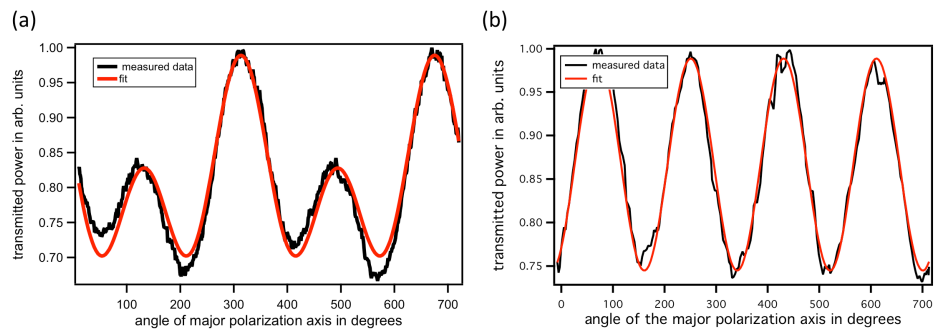

Fig. 4. Data from the polarimetry measurements. (a) Achromatic HWP. The data were fitted by a sum of two weighted $\cos^2$-functions. The data show a clear deviation from the expected behavior. Ideally, the peaks have the same height and are separated by 180 degrees from each other. (b) Superachromatic HWP. The peaks have the expected height and periodicity.

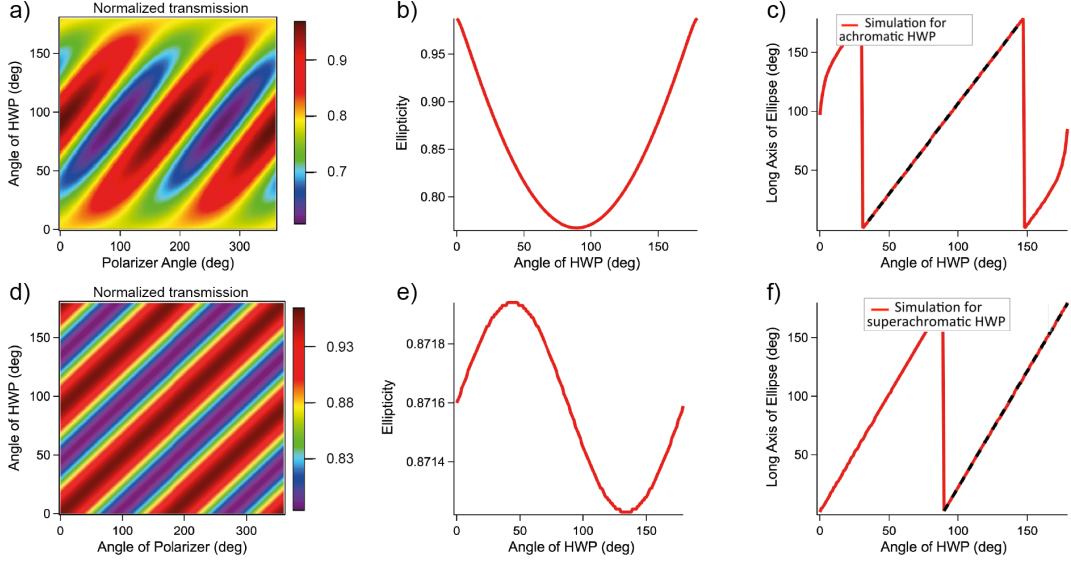

Fig. 5. Simulations of the pulse propagation through the achromatic (upper row, a-c) and superachromatic (lower row, d-f) waveplates. The plots show the transmission through a polarizer as a function of the orientations of the polarizer and the HWP (a and d, the ellipticity as a function of the rotation angle of the HWP (b and e), and the dependence of the angle of the major axis of the polarization ellipse on the rotation of the HWP (c and f). The results demonstrate how crucial the proper choice of waveplate is.

## 4. Data processing and results

### 4.1 Tomographic reconstruction

In principle the VMIS only allows to record two-dimensional projections of the actual three-dimensional electron distribution in the momentum space. In its typical geometry, a VMIS does not allow to record projections in the polarization plane, i.e. the plane of interest in angular streaking experiments. This matter can be overcome with tomographic reconstruction. A widely used, fast and efficient method for tomographic reconstruction is the filtered back-projection algorithm [21]. The combination of tomographic reconstruction with VMI was pioneered [10, 11] and successfully applied [12, 13] by various research groups in experiments on atoms and molecules. The algorithm performs the slice-wise reconstruction along the laser propagation direction on the basis of a Fourier transform. The reconstructed momentum distribution for each slice in the direction of beam propagation can be expressed as follows

$$f(p_x, p_y) = \int_{-\infty}^{+\infty} d\omega \int_0^{\pi} d\theta P(\omega, \theta) |\omega| e^{i 2\pi \omega t} \tag{1}$$

where $\theta$ denotes the projection angle, ω the frequency space coordinate and $P(\omega, \theta)$ the Fourier transform of the corresponding projection [21]. The momentum coordinate $p_y$ points along the major axis and the momentum coordinate $p_x$ along the minor axis of the polarization ellipse. The parameter $t = p_x \cos\theta + p_y \sin\theta$ denotes the coordinate along the projection plane. The reconstruction is performed slicewise along the beam propagation direction. In order to reduce artifacts from the reconstruction and to filter out high-frequency noise, a filter function with an appropriate cut-off frequency can be applied during the reconstruction. A variety of filters were designed for this purpose. In our case, a Hann-filter proved to be an excellent choice for the reconstruction of the full three-dimensional electron momentum distribution. If one is only interested in the projection onto the polarization plane, it is possible to project the electron images onto this plane prior to the reconstruction. For the analysis of the streaking angle we chose the latter method for numerical efficiency and without any significant effect on the extracted angle. Since the detector showed slightly non-uniform detection efficiency, the images were symmetrized on the beam propagation axis. As an angular step size, two degrees turned out to be an excellent choice for our purposes.

*4.2 Analysis*

For consistency, the same analysis approach as for earlier measurements with the COLTRIMS was chosen. The radially integrated angular distribution of the electrons was calculated from the reconstructed photoelectron distributions that were projected along the beam propagation direction, i.e. the data were integrated along the beam propagation direction, and fitted with a double peak Gaussian function. The peak location determined in this way was used as the streaking angle. An example for a photoelectron angular distribution together with a fit is shown in the inset of Fig. 7 Potential ionization delay times manifest themselves in an offset from a theoretical curve assuming instantaneous ionization. The fitting procedure is justified since the same method was used in the simulations that are used as a reference for the time zero. In our case, classical trajectory Monte-Carlo simulations using the tunnel ionization in a field induced tunnel barrier in parabolic coordinates was used for helium [3]. The tunnel barrier is further modified by an induced dipole and Stark shift in other noble gases as described with the TIPIS (tunnel ionization in parabolic coordinates with induced dipole and Stark shift) model [3].

*4.3 Results and comparison with data from COLTRIMS*

A projection and an isosurface of the reconstructed electron momentum distribution are depicted in Fig. 6. The shapes of both plots correspond to the expectations from previous experiments and theoretical considerations. The projection shows a ring artifact at the outer region of the electron distribution. The artifact stems from the fact that the data are slightly cut at the outside by the edge of the charge-coupled device (CCD) camera and the detector. The ring artifact was excluded in the calculation of the angular distributions. The streaking angles were compared to the ones that were recorded with the COLTRIMS setup. Comparison between the data sets that are plotted together in Fig. 7 demonstrates that the data from the VMIS setup lie within the error bars of the COLTRIMS data taken from Ref. [7]. The error bars of the VMIS-data stem from a geometrical sum of the errors from various sources. The sum is composed of the fit of the angular distribution of the electrons, the deviation in the result of the polarimetric measurements before and after the data acquisition and error due to inaccuracy of laser beam alignment. The latter one was estimated to +/- 0.5 degree due to the alignment inaccuracy of the incoming beam onto the back-focusing mirror. Furthermore, an estimated error of +/- 0.5 degree based on construction tolerances was added. The consistency between the data from both setups confirms that our implementation of the attoclock with a VMIS works as expected, in particular considering that based on the measured data and simulations the streaking angle is expected to be flat in this intensity range.

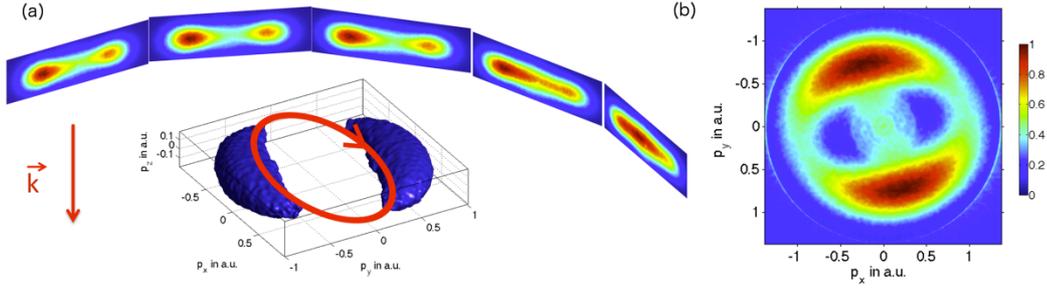

Fig. 6. Representations of a photoelectron momentum distribution recorded at an intensity of 0.2 PW/cm$^2$ and an ellipticity of 0.87. (a) Isosurface with original projections of the reconstructed photoelectron momentum distribution. The distribution was smoothened with a Gaussian spatial filter for a better visibility of the overall shape. The vector k marks the direction of the incoming light. (b) A projection along the beam propagation direction of the electron momentum distribution. The outer ring is an artifact from the reconstruction.

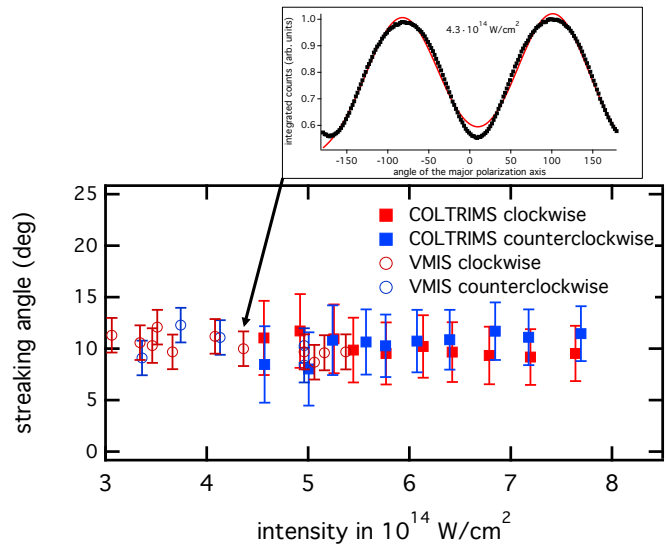

Fig. 7. Comparison of the streaking angles from the VMIS and the COLTRIMS setup as a function of the peak intensity of the ionizing laser pulse. The streaking angles show an excellent agreement within the error bars. The red points have been measured with clockwise elliptical polarization, the blue points with counterclockwise polarized pulses. The inset shows a typical angular distribution of the photoelectrons projected onto the polarization plane together with a double Gaussian fit. The black arrow indicates the data point that corresponds to the angular distribution.

## 5. Conclusion

We presented an alternative implementation of the attoclock with few-cycle pulses at a high repetition rate and count rate per pulse. The comparatively low technical complexity on the side of the vacuum chamber and the electronics eases the use of more complex targets, like e.g. molecules in gas phase. The high count and repetition rates and the correspondingly reduced acquisition time allows one to vary several parameters in the measurements while maintaining reasonable measurement times. In this work we varied the intensity of the driving laser. Another major advantage with respect to a COLTRIMS setup is the high momentum resolution that can be achieved in a VMI. However, this comes at the price of losing the ability to perform coincidence measurements. Few VMI setups have been reported to be capable of coincidence detection [22-24].

Here, we described the experimental details transferring the attoclock technique to a VMIS with few-cycle pulses. This approach fully benefits from high repetition and count rates. The essential electron momentum distribution was retrieved with tomographic reconstruction. Crucial elements are the waveplates, in particular the HWP that is rotated to record images for the set of angles that is required for the tomographic reconstruction. It was demonstrated in the experiment and supported by simulations that the usage of superachromatic waveplates is necessary for the broad bandwidth that is associated with few-cycle pulses. The continuous gas flow and the back-focusing geometry that allows high peak intensity and short pulses at a high target density calls for further applications in atomic, molecular and optical science with a variety of laser sources and targets.

**Acknowledgments**
This research was supported by the National Centres of Competence in Research Molecular Ultrafast Science and Technology (NCCR MUST), research instrument of the Swiss National Science Foundation (SNSF). This project has been co-financed under FP7 Marie Curie COFUND.